\documentclass[a4paper]{article}

\usepackage[pages=all, color=black, position={current page.south}, placement=bottom, scale=1, opacity=1, vshift=5mm]{background}
\SetBgContents{
	\tt This work is shared under a \href{https://creativecommons.org/licenses/by-nc-nd/4.0/}{CC BY-NC-ND 4.0 license} unless otherwise noted
}      

\usepackage[margin=1in]{geometry} 

\usepackage{amsmath}
\usepackage{amsthm}
\usepackage{amssymb}

\usepackage{booktabs}
\usepackage{array}
\usepackage{enumerate}

\usepackage[utf8]{inputenc}
\usepackage{hyperref}
\hypersetup{
	unicode,
	pdfauthor={Silva \textit{et al.}},
	pdftitle={ECKO: A semantic knowledge graph for personalized oncology},
}

\usepackage[sort&compress,numbers,square]{natbib}

\usepackage{graphicx, color}
\graphicspath{{fig/}}
\usepackage{todonotes}

\usepackage{algorithm, algpseudocode} 
\usepackage{mathrsfs} 

\usepackage{lipsum}

\title{ECKO: Explainable Clinical Knowledge for Oncology}
\author{Marta Contreiras Silva$^1$ \and Daniel Faria$^2$ \and Laura Balbi$^1$ \and Susana Nunes$^1$ \and Ana Filipa Rodrigues$^1$ \and Aleksander Palkowski$^{3,4,5}$ \and Michal Waleron$^{3,4,5}$ \and Emilia Daghir-Wojtkowiak$^3$ \and Ashwin Adrian Kallor$^3$ \and Christophe Battail$^6$\and Federico Maria Corazza$^7$\and Manuel Fiorelli$^8$ \and Armando Stellato$^8$ \and Javier Antonio Alfaro$^{3,4,5,9}$\and Fabio Massimo Zanzotto$^{10}$ \and Catia Pesquita$^1$}

\date{
	$^1$LASIGE, Faculdade de Ciências, Universidade de Lisboa, Portugal \\%
	$^2$INESC-ID, Instituto Superior Técnico, Universidade de Lisboa, Portugal \\%
    $^3$International Centre for Cancer Vaccine Science, University of Gdansk, ul. Kładki 24, Gdańsk 80-822, Poland \\
    $^4$The Riddell Centre for Cancer Immunotherapy, Arnie Charbonneau Cancer Institute, University of Calgary, HMRB 372, 3330 Hospital Drive NW, Calgary, Alberta, T2N 4N1, Canada \\
    $^5$Department of Biochemistry and Molecular Biology, Cumming School of Medicine, University of Calgary, HMRB 231, 3330 Hospital Drive NW, Calgary, Alberta, T2N 4N1, Canada \\
    $^6$ University Grenoble Alpes, IRIG, Laboratoire Biosciences et Bioingénierie pour la Santé, UA 13 Inserm-CEA-UGA, 38000 Grenoble, France\\
    $^7$Dstech s.r.l., Milan, Italy \\
    $^8$Department of Enterprise Engineering, Tor Vergata University of Rome, via del Politecnico 1, 00133 Roma RM, Italy\\
    $^9$School of Informatics, University of Edinburgh, Informatics Forum, 10 Crichton St, Newington, Edinburgh EH8 9AB \\
    $^{10}$Human-Centric ART, Tor Vergata University of Rome, via del Politecnico 1, 00133 Roma RM, Italy\\
}

\begin{document}
	\maketitle
	
	\begin{abstract}

Personalized oncology aims to tailor treatment strategies to the unique molecular and clinical profiles of individual patients, moving beyond the traditional paradigm of treating the disease not the patient. Achieving this vision requires the integration and interpretation of vast, heterogeneous biomedical data within a meaningful scientific framework. Knowledge graphs, structured according to biomedical ontologies, offer a powerful approach to contextualize and interconnect diverse datasets, enabling more precise and informed clinical decision-making.

We present ECKO (Explainable Clinical Knowledge for Oncology), a comprehensive knowledge graph that integrates 33 biomedical ontologies and aggregates data from multiple studies to create a unified resource optimized for data-driven clinical applications in oncology. Designed to support personalized drug recommendations, ECKO facilitates the identification of optimal therapeutic options by linking patient-specific molecular data to relevant pharmacological knowledge. It provides transparent, interpretable explanations for drug recommendations, fostering greater trust and understanding among clinicians and researchers. This resource represents a significant advancement toward explainable, scalable, and clinically actionable personalized medicine in oncology, with potential applications in biomarker discovery, treatment optimization, and translational research.
	\end{abstract}

\section{Background and Summary}
The successful application of Artificial Intelligence (AI) to personalized medicine depends on the integration of vast amounts of data across different biomedical domains~\cite{chin2011cancer}. When this integration is supported by knowledge-based techniques, such as ontologies and knowledge graphs, AI predictions have been shown to be both more accurate and relevant~\cite{althubaiti2019ontology, jimenez2020drug, dragoni2022knowledge}, and their explanations to be more human-understandable~\cite{Nunes} -- an essential step toward addressing the black-box nature of many AI models, which remains a major concern for healthcare professionals~\cite{Xu}. 

Despite recent efforts in integrating biomedical data from multiple repositories into knowledge graphs for personalized medicine~\cite{morris2023scalable, primekg, quan2023aimedgraph},  these fall short not only due to their inability to properly model patient and sample data, but also because they fail to leverage the rich semantics afforded by biomedical ontologies. Ontologies, by providing rich models of the entities in a domain and the relations between them~\cite{silva2022ontologies}, are crucial to identify biologically relevant connections within and between datasets. Moreover, they are fundamental to placing AI predictions in the context of biomedical knowledge, which is crucial to support research and clinical practice that relies on a holistic understanding of each patient’s unique characteristics and medical history. These aspects are especially critical for personalized cancer treatment, where integrating heterogeneous data and ensuring explainability and contextualization of AI predictions can directly influence diagnostic accuracy, treatment selection, and patient outcomes.

\begin{figure}[h!]
    \centering
    \includegraphics[width=0.9\linewidth]{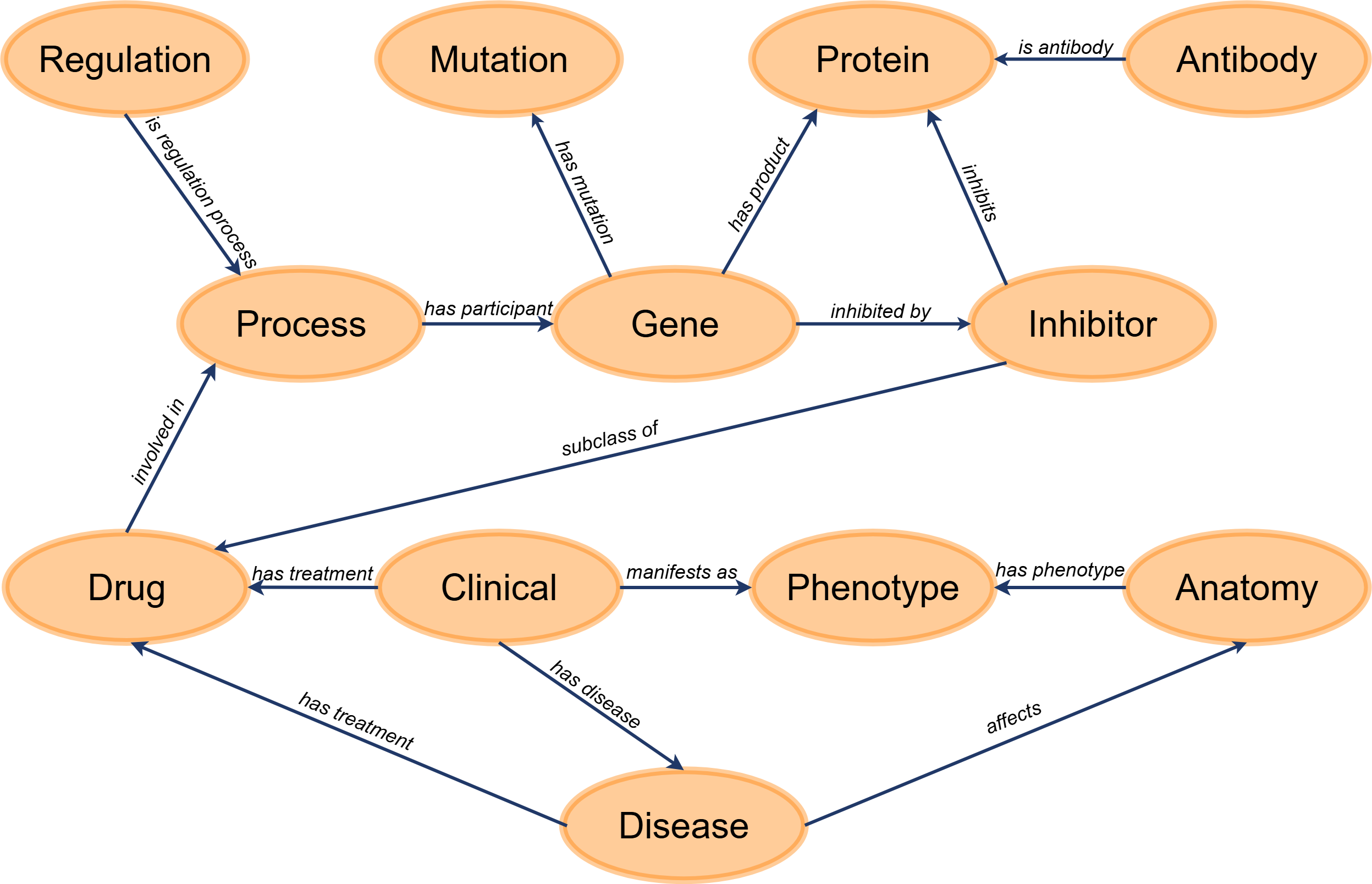}
    \caption{High-level schema of ECKO. This diagram represents the types of entities present in the KG, the relations established between them, and the ontologies that describe them.}
    \label{fig:kg}
\end{figure}

We present ECKO (Explainable Clinical Knowledge for Oncology), a Knowledge Graph for personalized oncology that integrates 33 biomedical ontologies and 74 datasets. The KG prioritizes a very rich ontological component to support scientific contextualization and explainability (Figure~\ref{fig:kg}). An ontology-rich KG is capable of supporting not only logical reasoning but also providing biologically relevant paths that can be leveraged into explanations, assisting healthcare professionals and researchers in their understanding of the related concepts. The ontological layer of the KG has 1,418,963 concepts and 2,736 types of edges to potentially link entities. Data entries were annotated by ontological concepts totaling 295,964,112 links, and ontologies were interconnected between themselves through 193,503 simple equivalences and 136,287 complex “related to” relationships. The KG has been evaluated in drug recommendation for oncology therapy and in generating explanations for predictions of gene-drug pairs.

While the current version of the KG has been tailored to oncological treatment, the network of ontologies enables its extension with datasets from other domains, ensuring its reusability. Moreover, the KG is constructed entirely with open-access data and ontologies, ensuring that there are no obstacles to its sharing and use.

\section{Methods}
The first step in constructing ECKO was selecting the ontologies that better represented the biomedical domain and ensured proper coverage, followed by choosing the publicly available datasets to be included. After identifying significant gaps in the representation of transcriptomic and immunopeptidomics data within the selected ontologies, we engaged in multiple expert consultations to develop two semantic data models that address these limitations. These models were then formalized into a new ontology to ensure comprehensive coverage in the KG. To ensure a seamless integration of all ontologies, we employed ontology alignment techniques to link equivalent entities between all ontologies efficiently. We further enriched these links with complex mappings capturing more complex relations between ontology entities. Domain datasets were selected and imbued into the context provided by the selected ontologies through data annotation, establishing links between specific sets of data points and biomedical concepts. \subsection{Ontologies and Data Sources}
\subsubsection{Ontologies}
Ontology selection is a crucial step in the construction of the KG as they must encompass all datasets that will be added. A list of data resources was provided by experts which was analyzed for their usage of ontologies. Afterwards, an extensive manual search in BioPortal~\cite{whetzel2011bioportal} – one of the main repositories of biomedical ontologies – was conducted to find additional ontologies that afford further coverage of the data domains. The initial list of ontologies was ranked according to relevance and quality, excluding ones that rated low in both.

As outlined in Table~\ref{tab:ontos}, the selected 33 biomedical ontologies cover a variety of domains, such as clinical features, genomics, drug side effects, clinical trials, and biological features, ensuring that they cover all necessary entities as shown in Figure~\ref{fig:kg}. 
The ontologies were incorporated into the KG without any processing, ensuring that all hierarchical and logical relations are included in the final graph.
To obtain the ontology files, OBO Foundry~\cite{smith2007obo} was prioritized, followed by the ontology’s own website or repository, and finally BioPortal, with a preference for the OWL format whenever available. The ontologies were collected in November 2021. 

\begin{table}
\centering
\footnotesize
\caption{Ontologies used in the Knowledge Graph, and their respective acronym, name, domains, number of classes, and reference.}
\label{tab:ontos}
\begin{tabular}{@{}m{1cm}m{4cm}m{4cm}lm{4cm}@{}}
\toprule
\textbf{Acronym} & \multicolumn{1}{c}{\textbf{Ontology}} & \multicolumn{1}{c}{\textbf{Domains}} & \multicolumn{1}{c}{\textbf{Classes}} & \textbf{Reference} \\ \midrule
ACGT-MO & Cancer Research and Management ACGT Master Ontology & Clinical feature, sample status & 1769 & Brochhausen \textit{et al.}\cite{acgt} \\
ATC & Anatomical Therapeutic Chemical Classification & Drug & 6567 & \url{https://atcddd.fhi.no/} \\
BFO & Basic Formal Ontology & Properties & 36 & Spear \textit{et al.}\cite{bfo} \\
CCTOO & Cancer Care: Treatment Outcome Ontology & Response to treatment, drug screening & 1133 & Lin \textit{et al.}\cite{cctoo} \\
ChEBI & Chemical Entities of Biological Interest Ontology & Metabolic, drug & 171058 & Hastings \textit{et al.}\cite{Hastings_Owen_Dekker_Ennis_Kale_Muthukrishnan_Turner_Swainston_Mendes_Steinbeck_2016} \\
CL & Cell Ontology & Cellular & 10984 & Diehl \textit{et al.}\cite{cl} \\
CLO & Cell Line Ontology & Cell line & 44873 & Sarntivijai \textit{et al.}\cite{clo} \\
CMO & Clinical Measurement Ontology & Clinical feature, sample status & 3054 & Shimoyama \textit{et al.}\cite{cmo} \\
DCM & DICOM Controlled Terminology & Histological images & 4561 & \url{https://dicom.nema.org/medical/dicom/current/output/chtml/part16/chapter\_D.html} \\
DOID & Human Disease Ontology & Clinical feature & 17642 & Schriml \textit{et al.}\cite{doid} \\
DTO & Drug Target Ontology & Drug target interactions & 10075 & Lin \textit{et al.}\cite{dto} \\
EFO & Experimental Factor Ontology & Experimental & 28816 & Malone \textit{et al.}\cite{efo} \\
FMA & Foundational Model of Anatomy & Anatomical data & 78977 & Cook \textit{et al.}\cite{fma} \\
GENO & Genotype Ontology & Genomic & 425 & \url{https://github.com/monarch-initiative/GENO-ontology/} \\
GO & Gene Ontology & Genomic, biological pathway & 50713 & The Gene Ontology Consortium \textit{et al.}\cite{go} \\
HCPCS & Healthcare Common Procedure Coding System & Clinical feature, drug sampling & 7094 & \url{https://www.cms.gov/medicare/coding-billing/healthcare-common-procedure-system} \\
HGNC & HUGO Gene Nomenclature & Genomic & 32917 & Seal \textit{et al.}\cite{hgnc} \\
HP & Human Phenotype Ontology & Biological feature & 27482 & Gargano \textit{et al.}\cite{hp} \\
ICDO & International Classification of Diseases Ontology & Clinical feature & 1313 & \url{https://icd.who.int/en} \\
ImPO & Immunopeptidomics Ontology & Immunopeptidomics, experimental & 68 & Faria \textit{et al.}\cite{impo} \\
LOINC & Logical Observation Identifier Names and Codes & Clinical feature & 268552 & McDonald \textit{et al.}\cite{mcdonald2003loinc} \\
MONDO & Mondo Disease Ontology & Clinical feature & 43735 & Vasilevsky \textit{et al.}\cite{mondo} \\
NCIT & National Cancer Institute Thesaurus & Biological feature, clinical feature & 166884 & Hartel \textit{et al.}\cite{hartel2005modeling} \\
OAE & Ontology of Adverse Effects & Drug side effect, response to treatment & 5762 & He \textit{et al.}\cite{oae} \\
OMIM & Online Mendelian Inheritance in Man & Biological feature & 97261 & \url{https://www.omim.org/} \\
OPMI & Ontology of Precision Medicine and Investigation & Clinical feature, clinical trial & 2939 & He \textit{et al.}\cite{opmi} \\
ORDO & Orphanet Rare Disease Ontology & Clinical feature & 14886 & \url{https://www.orphadata.com/ordo/} \\
PDQ & Physician Data Query & Clinical feature, drug screening & 13452 & Hubbard \textit{et al.}\cite{hubbard1986physician} \\
PMAPP-PMO & PMO Precision Medicine Ontology & Genomic, clinical feature, clinical trial, sampling & 76154 & Hou \textit{et al.}\cite{hou2020pmo} \\
RO & Relation Ontology & Properties & 58 & \url{https://github.com/oborel/obo-relations/} \\
SO & Sequence Ontology & Genomic, transcriptomic & 2707 & Eilbeck \textit{et al.}\cite{so} \\
UBERON & Uber-anatomy Ontology & Anatomical data & 26368 & Mungall \textit{et al.}\cite{uberon} \\ 
\bottomrule
\end{tabular}
\end{table}

\subsubsection{Data Sources}
The datasets included belong to two main domains: immunopeptidomics and transcriptomics.
The immunopeptidomics data originate from the work of Waleron \textit{et al.}~\cite{Waleron} that produced the CAnceR iMmunopEptidogeNomics (CARMEN) database to research MHC Class I binding promiscuity for vaccine discovery. The data, gathered from 72 datasets, was determined through different mass-spectrometry methods for a total of 2323 samples derived from cell culture, tissue, and mixed sources corresponding to healthy or tumoral conditions.

The transcriptomics data resulted from the combination of two cohort datasets, the Braun~\cite{braun2020interplay} and the Motzer~\cite{motzer2020avelumab} datasets. 
The Braun dataset comprises multi-modal molecular and immunological profiles from nearly 600 advanced clear-cell renal cell cancer patients treated with anti–PD-1 (programmed cell death receptor) therapy. The patient profiles were built from the linking of whole-exome sequencing, transcriptome profiling, and multiplex immunofluorescence experimental readouts to clinical outcomes, including objective response rates and progression-free survival.
The Motzer dataset resulted from the \textit{JAVELIN Renal 101} trial with nearly 900 treatment-naive advanced renal cell cancer patients treated either with sunitinib or with the drug combination of avelumab and axitinib. The dataset integrated patients' sequencing data outputs with their progression-free survival and response outcomes to enable the identification of biomarkers predictive of therapeutic benefit.

\subsection{Semantic Data Models}
Through the analysis of the datasets, it was noted that the experimental description of protocols and data was lacking in appropriate granularity. A new ontology, the ImmunoPeptidomics Ontology~\cite{impo}, was developed to fill in these gaps in experimental data annotation.

\subsubsection{The Immunopeptidomics Ontology}
The ImmunoPeptidomics Ontology (ImPO) was developed with the help of experts in the domain, through an extensive analysis of existing immunopeptidomics datasets in order to model their classes and properties to suit real-life data. As an approximation of the final ontology, an Entity-Relationship (ER) model was constructed from the existing semantics, which was successively updated following multiple discussions with experts, until a finalized model was obtained. This ER model was formalized as an ontology using the OWL format, totaling 47 classes, 36 object properties, and 39 data properties.

\subsubsection{The Transcriptomics Ontology}
Transcriptomics, while distinct, has a high degree of overlap with immunopeptidomics in its experimental design. As such, any gaps in the semantic model for this specific domain were filled by extending ImPO with new classes and properties. The final ImPO ontology has 68 classes, 39 object properties, and 81 data properties, and is available at \url{https://github.com/liseda-lab/ImPO}.

\subsection{Ontology Alignment}
Equivalences were established between classes in the ontologies, effectively aligning them, and ensuring they form an interconnected semantic layer that supports the KG. The number of ontologies (and respective entities) involved in this task presented a challenge in scalability, while their granularity and differing points of view presented a challenge in their conciliation. Simple 1:1 equivalences were found between all ontologies, while more indirect “related to” links were established by finding complex 1:n multi-ontology constructs.

\begin{figure}[h!]
    \centering
    \includegraphics[width=0.6\linewidth]{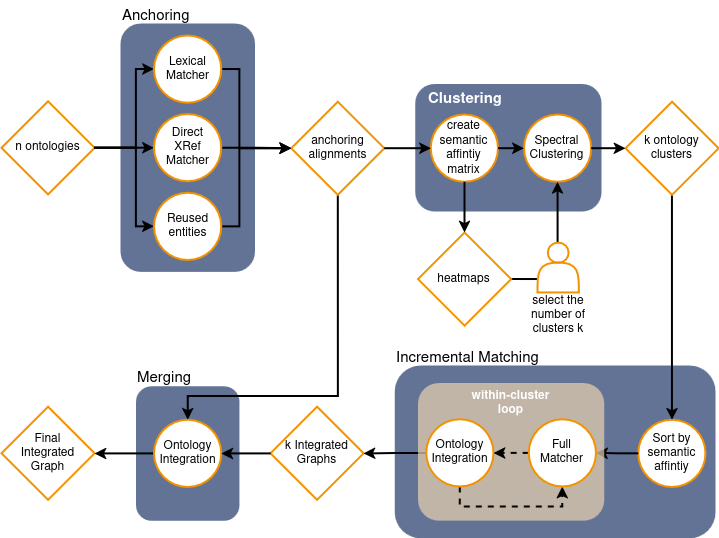}
    \caption{General overview of the Holistic Ontology Matching (HOM) methodology adapted from Silva \textit{et al.}\cite{hom}.}
    \label{fig:hom}
\end{figure}

\subsubsection{Semantic Data Model Mappings}
The ImmunoPeptidomics Ontology was aligned to the other 32 biomedical ontologies using AgreementMakerLight (AML)\cite{aml} through its default pipeline. This process was run as a series of pairwise matching tasks of ImPO to each ontology. The mappings found were manually validated by a domain expert to ensure that all mappings were scientifically accurate. This process yielded a total of 185 final mappings that were both encoded into the ontology as cross-references and also saved to a single alignment file.

\subsubsection{Holistic Ontology Matching}
The scalability issue of matching 32 ontologies was addressed using a holistic ontology matching approach by Silva \textit{et al.}~\cite{hom} which leverages the overlap between biomedical ontologies to divide them into clusters, which are then aligned following an incremental strategy, reducing the number of redundant actions. An overview of this strategy can be found in Figure~\ref{fig:hom}.

\begin{figure}[h!]
    \centering
    \includegraphics[width=0.7\linewidth]{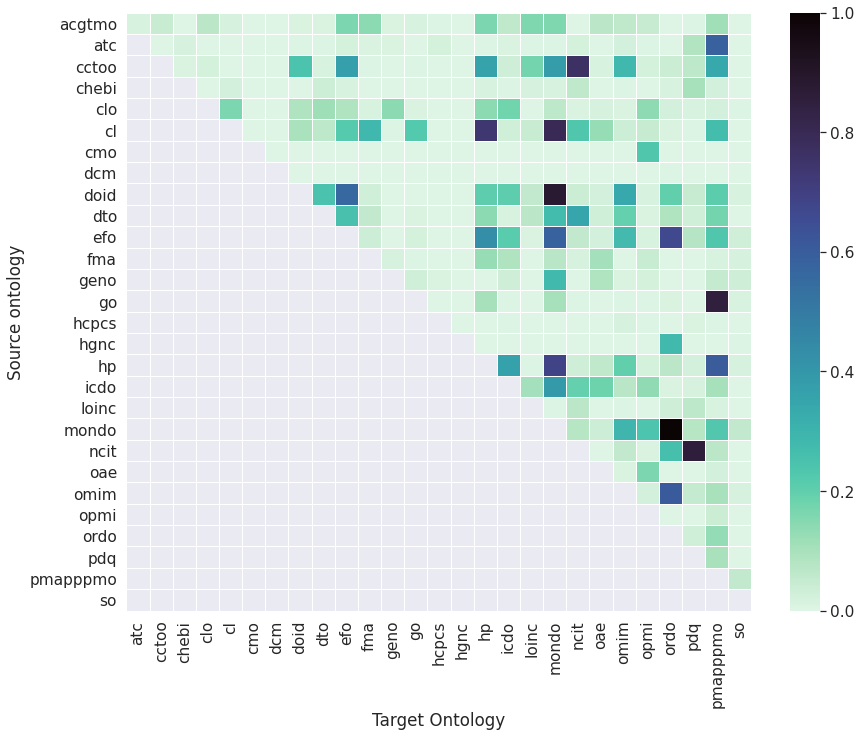}
    \caption{Anchoring heatmap for Holistic Ontology Matching from Silva \textit{et al.}\cite{hom}.}
    \label{fig:matrix}
\end{figure}

To create the clusters, an initial anchoring step was performed that calculates the overlap between all pairs of ontologies, using fast and high-confidence algorithms. This overlap is calculated as the fraction of classes of the smallest ontology of each pair that have the same URI, direct cross-references, shared cross-references, overlapping logical definitions, or equivalent labels or synonyms to classes in the largest ontology of the pair. This overlap can be visualized as an affinity matrix in Figure~\ref{fig:matrix}.

The affinity matrix was used to divide the ontologies into four clusters using spectral clustering, under the assumption that a higher level of overlap translates to a higher likelihood that the ontologies belong to the same subdomain. Within each cluster, the ontologies are then aligned incrementally, meaning a first ontology is aligned to a second ontology, and the resulting alignment is then aligned to a third ontology, and so forth in descending order of overlap value. The result of this incremental process is a single alignment that avoids redundant mappings and consequently decreases the overall runtime of the process.
The matching used the AML system and its automatic pipeline with default configurations, and yielded 193,503 mappings.

\subsubsection{Complex Multi-Ontology Matching}
The biomedical domain can be described with far more granularity than what can be accurately captured by 1:1 equivalences. For example, some biomedical ontologies contain logical definitions, which are 1:n complex mappings where a single concept is mapped to an expression composed of multiple concepts (e.g., \textit{decreased circulating cortisol level} $\equiv$ (\textit{has part} SOME (\textit{decreased amount} AND (\textit{inheres in} SOME (\textit{cortisol} AND (\textit{part of} SOME \textit{blood}))) AND (\textit{has modifier} SOME \textit{abnormal})))). Finding and adding such complex equivalencies to the KG establishes additional links that codify relatedness between concepts.

\begin{figure}[h!]
    \centering
    \includegraphics[width=\linewidth]{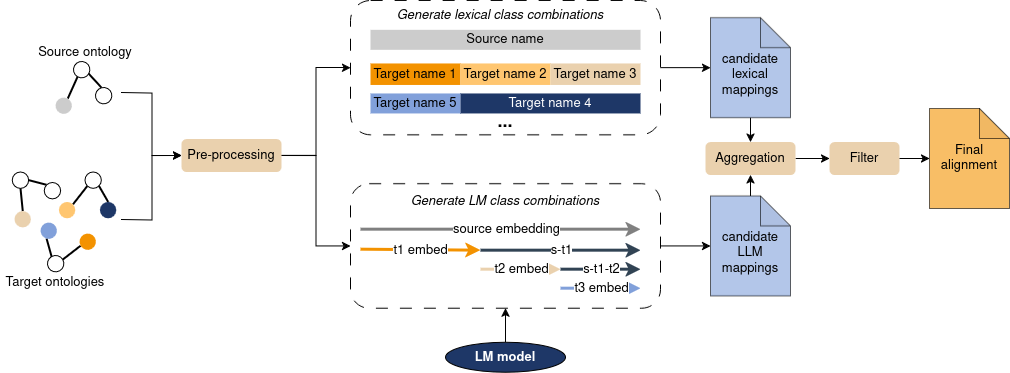}
    \caption{General overview of CMOM methodology adapted from Silva \textit{et al.}~\cite{cmom-rs}}
    \label{fig:cmom}
\end{figure}

To find these links, we used a complex multi-ontology matching approach by Silva et al.~\cite{cmom-rs} that combines a lexical strategy with a language model (LM)-based strategy and returns sets of entities that can be combined into a logical expression. A general overview of this method can be seen in Figure~\ref{fig:cmom}.

Only three ontologies were used as sources: CL, HP, and UBERON, due to their relevance and their incorporation of complex concepts that overlap with the domains of other ontologies. Each of these was matched against a subset of target ontologies which can be seen in Table~\ref{tab:cmom_srcs}.

\begin{figure}
    \centering
    \includegraphics[width=0.8\linewidth]{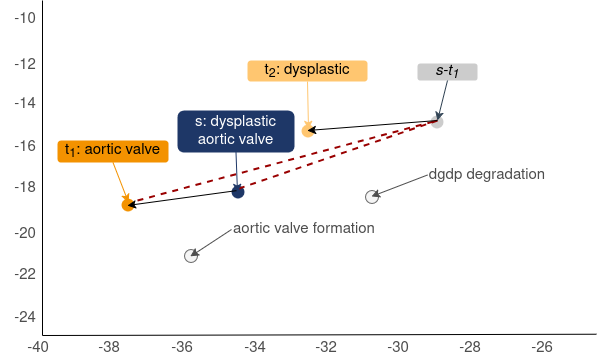}
    \caption{Visualization of the first two steps of the construction of a complex mapping in a 2D space using a geometric operation.}
    \label{fig:tsne}
\end{figure}

An initial pre-processing of the lexical information of the ontologies, ensures that all names and synonyms are normalized. The lexical approach selects classes by initially filtering for all target names that share at least one word with the source name, followed by a recursive approach that finds all possible combinations of these target names that do not overlap and provide full coverage of the source name. The LM-based class selection relies on recursive subtraction of the generated sentence embeddings, as shown in Figure~\ref{fig:tsne}, whereas the most similar target embedding to the source embedding is subtracted from it, and this new vector is used to find the next most similar target embedding, and so on. The embeddings were generated with the sentence-BERT model~\cite{bert} (which can be replaced by any other similar model), and cosine similarity was used to compute the similarity between the embeddings.

In order to aggregate and filter the generated candidates, a greedy heuristic was employed to select mappings sorted by descending order, producing a (near) 1:1 alignment, where equal-value mappings are all returned. 
Complex mappings were incorporated into the KG by establishing links of “related to” between the sets of target classes and each simple source class. Of the 33 biomedical ontologies, only HP has pre-existing complex mappings totaling 35,800 links to 6,230 source classes. Using the CMOM strategy a further 100,487 links were found to 21,759 source classes.

\renewcommand{\arraystretch}{2}
\begin{table}[h!]
\footnotesize
\caption{Subsets of target ontologies for each of the source ontologies used in Complex Multi-Ontology Matching. The abbreviations used are outlined in Table~\ref{tab:ontos}.}
    \centering
    \begin{tabular}{m{5cm}|m{10cm}}
        \textbf{Source Ontology} & \textbf{Target Ontologies} \\
        \midrule
        Cell Ontology & ACGT, CLO, DCM, EFO, HP, ImPO, LOINC, NCIT, PMO \\
        Human Phenotype Ontology & ChEBI, CL, DOID, DTO, GENO, GO, ICDO, ImPO, LOINC, MONDO, NCIT, OMIM, ORDO, PATO, PMO, UBERON \\
        Uber-anatomy Ontology & ACGT, ATC, CL, CLO, CMO, DCM, DOID, FMA, HCPCS, HP, ICDO, ImPO, LOINC, MONDO, NCIT, OMIM, OPMI, ORDO, PMO, SO \\
    \end{tabular}
    \label{tab:cmom_srcs}
\end{table}

\subsection{Data Integration}\label{sec:data_int}
The ImPO ontology provides a standardized form of domain terminology and semantics and therefore serves as the common ground for all other ontologies that compose the KG. As such, the multi-omics experimental and analysis data were integrated with the latest version of the ImPO using the RDFLib package to model data in the RDF scheme. This process began by defining the mapping rules for each text data file, annotating the files’ column names to the ontology's entities, and finally creating RDF triples to capture the mapping rules for and between the files. The resulting instantiated KG contains over 112M entities and 295M statements.

Part of the immunopeptidomics data (genes, drugs) was also mapped directly to a few selected data sources, totaling 977,909 data mappings. Of those, 360,287 are direct mappings to four core ontologies – GO, ChEBI, HGNC and HP –, while the remaining 617,622 link to external databases covering genome annotation (Ensembl~\cite{ensembl}, KEGG~\cite{kegg}), sequence records (EMBL~\cite{embl}), structural information (AlphaFoldDB~\cite{alphafold}), proteomics (PeptideAtlas~\cite{peptideatlas}, ProteomicsDB~\cite{proteomicsdb}) and phylogenomics data (OMA~\cite{oma}), protein domains (Pfam~\cite{pfam}, SUPFAM~\cite{supfam}, PROSITE~\cite{prosite}, PANTHER~\cite{panther}, InterPro~\cite{interpro}, Gene3D~\cite{gene3d}), interaction networks (STRING~\cite{stringdb}), enzymatic pathways (Reactome~\cite{reactome}) and organism-specific resources (OpenTargets~\cite{opentargets}, OMIM~\cite{omim}, HPA~\cite{hpa}, GeneCards~\cite{genecards}).

\section{Data Availability}
The knowledge graph is available at \url{https://doi.org/10.5281/zenodo.15789542}. It contains 33 biomedical ontologies (5 in Turtle format and 28 in OWL format) covering various domains as shown in Table~\ref{tab:ontos}. It contains both immunopeptidomics and transcriptomics datasets, which were incorporated using the data integration methods described in Section~\ref{sec:data_int} and are shared in the form of 27 proteogenomics OWL files and 17 transcriptomics OWL files, which were mapped to ImPO, and 5 OWL files that map to other ontologies. The simple equivalences are shared as OWL alignments split into ontology pairs, totaling 496 files. The complex mappings are also shared as TSVs, split into two files: triples originating from the ontologies directly and triples from the alignments found using CMOM-RS.

\section{Technical Validation}

A summary of the statistics of ECKO can be found in Table~\ref{tab:stats}.

\begin{table}[h!]
\centering
\footnotesize
\caption{Statistics for ECKO.}
\label{tab:stats}
\begin{tabular}{@{}m{3cm}m{2cm}@{}}
\toprule
\textbf{Datasets} & 74 \\
\textbf{Ontologies} & 33 \\
\textbf{Classes} & 1,418,963 \\
\textbf{Properties} & 2,736 \\
\textbf{Instances} & 112,577,730 \\
\textbf{Data annotations} & 295,964,112 \\
\textbf{Simple Mappings} & 193,503 \\
\textbf{Complex Mappings}& 136,287 \\
\bottomrule
\end{tabular}
\end{table}

\subsection{Use Case: Drug Recommendations for Renal Cancer}

ECKO was validated in the context of the "KATY – Knowledge At the Tip of Your fingers: Clinical Knowledge for Humanity" project~\footnote{https://katy-project.eu/}. KATY aimed to generate an AI system for personalized oncology for predicting and explaining cancer treatment outcomes, building trust in clinicians. It targeted clear cell Renal Cell Carcinoma (ccRCC) and proposed to leverage publicly available data and ontologies.

Publicly available data generated using different -omics technologies (genomic, transcriptomic, proteomic data) are challenging to mine effectively due to their heterogeneity, complex interrelationships, structural complexity, noise, and sparsity. Efficiently extracting insights from such data requires a machine learning system capable of learning from incomplete information and iteratively acquiring new knowledge. 

\begin{figure}[h!]
    \centering
    \includegraphics[width=0.7\linewidth]{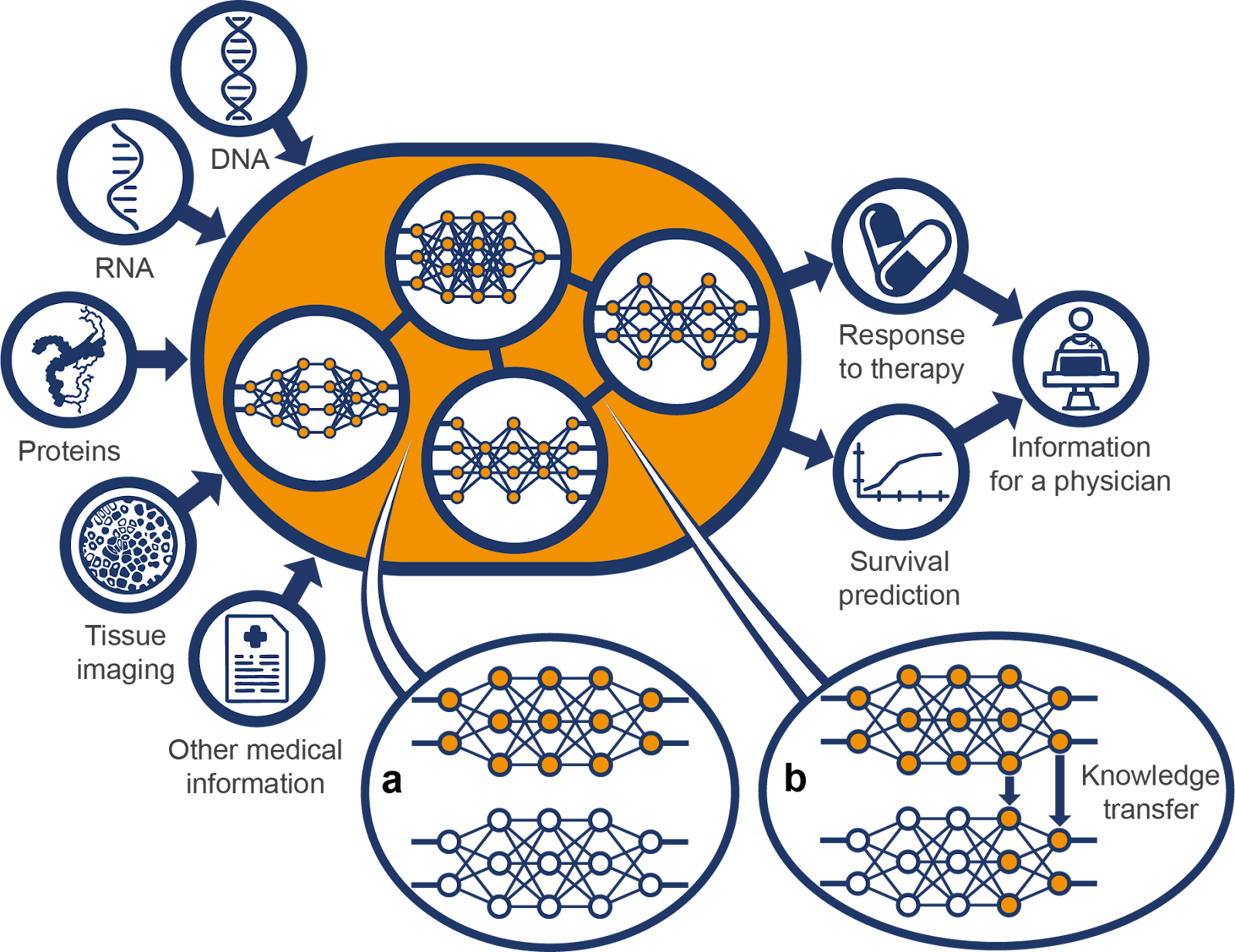}
    \caption{General representation of the KATY Holistic Neural Network model (KATYHoNN) from Daghir-Wojtkowiak \textit{et al.}~\cite{daghir2025leveraging}. (a) multi-task learning, (b) single task transfer learning.}
    \label{fig:honn}
\end{figure}

To tackle these challenges, the KATY project developed a Holistic Neural Network (KATYHoNN) (overview in Fig.~\ref{fig:honn}), which consists of several sub-networks, each taking as input specific omics or clinical data. 
The KATYHoNN model's input leverages publicly available datasets for ccRCC patients (e.g., RNA-Seq data, histological data) and data from clinical trials evaluating the efficiency of therapies. The heart of the model is composed of individual sub-networks for which the input is available omics data or clinical patient data. The sub-networks can be trained either on: (i) a singular specific task (e.g., either genomics, transcriptomics, or proteomics) via transfer learning, (ii) all tasks (i.e., the prediction of the best treatment recommendation based on molecular and cellular characteristics, as well as clinical and biological metadata from patients and samples) via multi-task learning to compile the general network, or (iii) be used to fine-tune a model adjusted to a new task by leveraging knowledge from a related domain (sequential transfer learning). The result of multi-task learning is the prediction of response-to-therapy, while the results from transfer learning are the prediction of other features not directly related to treatment choice, but relevant to it.

This divide-and-conquer approach contributes to a final model that is both easier to train and more accurate, as each sub-network is trained for a specific task and can be easily fine-tuned to handle changes. It is possible to both add new components within the KATYHoNN model and modify the existing ones by deleting or linking them. This training strategy also mitigates the missing data problem, where there may be sufficient data for a specific task but insufficient data for all tasks combined. In case of insufficient data for model training, data from different cancer types can be used to pre-train specific sub-networks, which can then be transferred to the ccRCC model.

To be clinically useful, AI predictions must be both accurate and understandable. The final step in the KATYHoNN framework is the identification of the most relevant genes for a given drug recommendation using SHAP~\cite{shapley} and LIME~\cite{lime}. However, knowing which genes were most relevant to support a drug recommendation does not necessarily provide sufficient biological context to support a clinician's understanding of the scientific validity of the AI outcome. ECKO was thus validated in providing explanations of the scientific validity of the KATYHoNN predictions based on the most relevant genes. To accomplish this, ECKO was deployed as part of the KATY platform.

\subsection{Deployment}
The KATY platform consists of a multi-container Docker application on an Amazon EC2 instance secured using AWS Identity and Access Management roles and security groups to control network access and permissions. ECKO was deployed in a GraphDB~\cite{graphdb} instance (version 10.5.1) without transitive closure. Explanations are visualized using D3.js.

\begin{figure}[h!]
    \centering
    \includegraphics[width=0.7\linewidth]{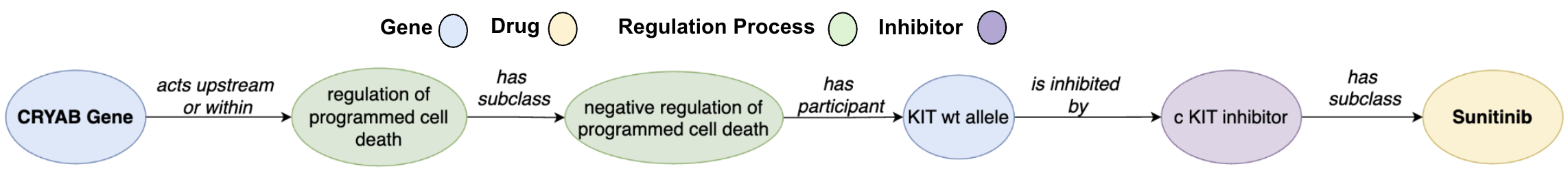}
    \caption{Drug recommendation path between a patient expressed gene, the CRYAB gene,  and the drug Sunitinib.}
    \label{fig:exp}
\end{figure}

\subsection{Generation of Explanations}
To validate ECKO, we generated explanations for the personalized drug recommendations predicted by the KATYHoNN, drawing inspiration from the REx framework~\cite{Nunes}. REx formulates scientific explanations as KG paths that connect two entities, in our case, the relevant gene and the recommended drug. These paths prioritize fidelity, which reflects whether the explanation aligns correctly with the prediction, and relevance, measured by the informativeness of the nodes in the paths. 

While REx employed reinforcement learning to find explanatory paths, given the scale of our KG, we opted for an approach with less computational cost. Instead of training a reinforcement learning agent, we implemented a k-shortest paths algorithm guided by the same fidelity and relevance criteria, repurposed as a ranking function rather than as a reward mechanism. This allowed us to efficiently explore meaningful explanatory paths, preserving the core principles of REx, while avoiding the high computational overhead typically associated with reinforcement learning in large-scale graph environments.

\begin{figure}[b!]
    \centering
    \includegraphics[width=0.6\linewidth]{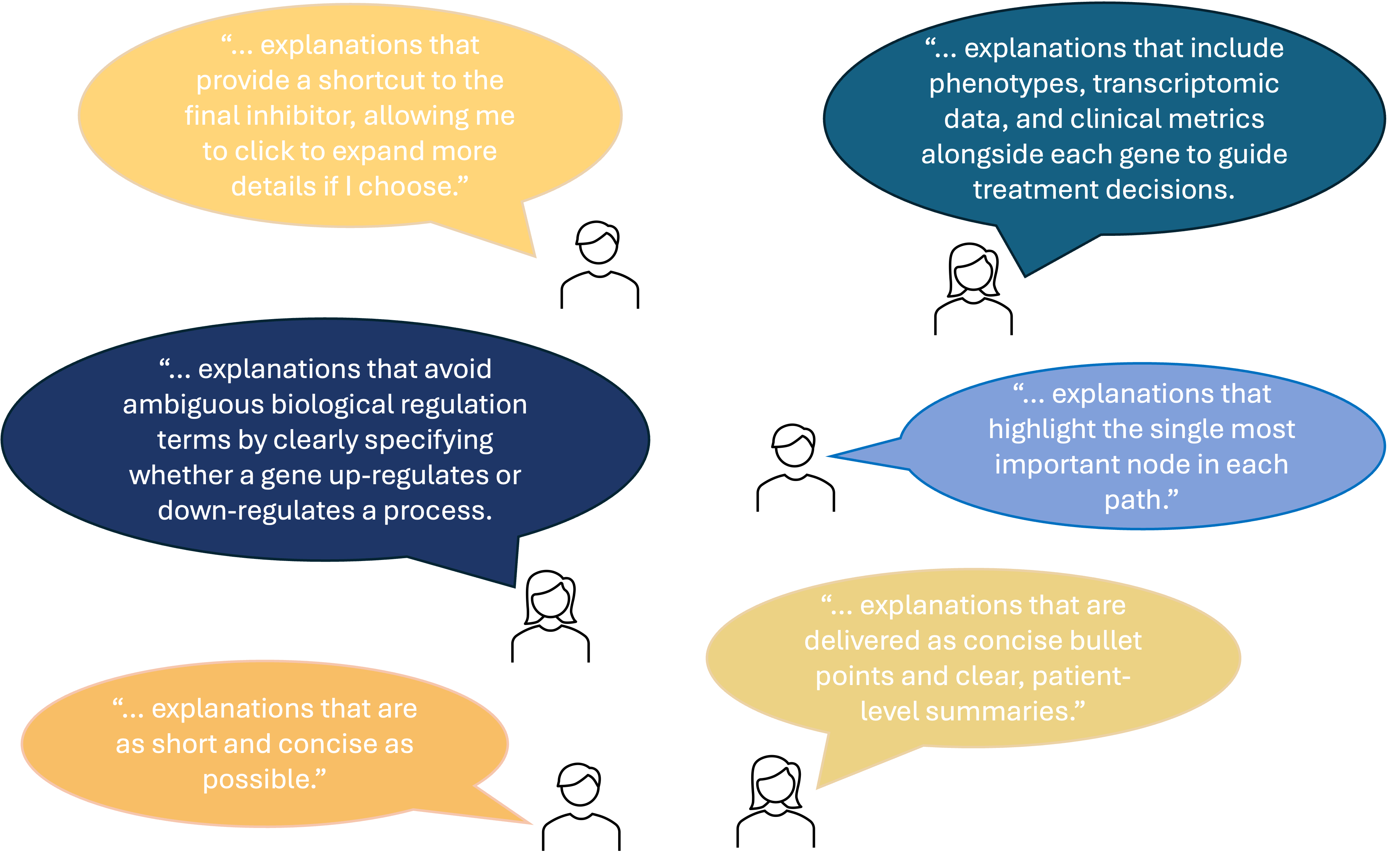}
    \caption{Representative comments from focus group, illustrating preferences for clarity, biological grounding, clinical relevance, and concise presentation of explanatory paths.}
    \label{fig:com}
\end{figure}

Figure~\ref{fig:exp} presents an illustrative example that links a gene expressed by the patient (the CRYAB gene), which was identified as relevant to a specific drug recommendation (the drug Sunitinib). The path provided by our approach offers insight into the rationale behind the prediction by highlighting the most relevant explanatory path connecting these two entities. In this specific example, the path includes other entities such as regulation processes and inhibitors that allow the prediction to be anchored into biomedical context, increasing the understanding and trust of the clinicians and researchers in the AI outcome.

Six examples of explanatory paths were validated by six focus groups, gathering 26 specialists --- including clinicians, biomedical researchers, and bioinformaticians. Discussions and qualitative evaluations within the group highlighted a clear preference for explanatory paths that explicitly represent underlying biological phenomena, emphasizing the importance of ontological knowledge in supporting clinical and research decision-making (Figure~\ref{fig:com}).

ECKO was able to support explanations for all 296 test patient recommendations. For each patient, we were able to generate explanations for all of their most relevant genes.

\begin{figure}[h!]
    \centering
    \includegraphics[width=0.6\linewidth]{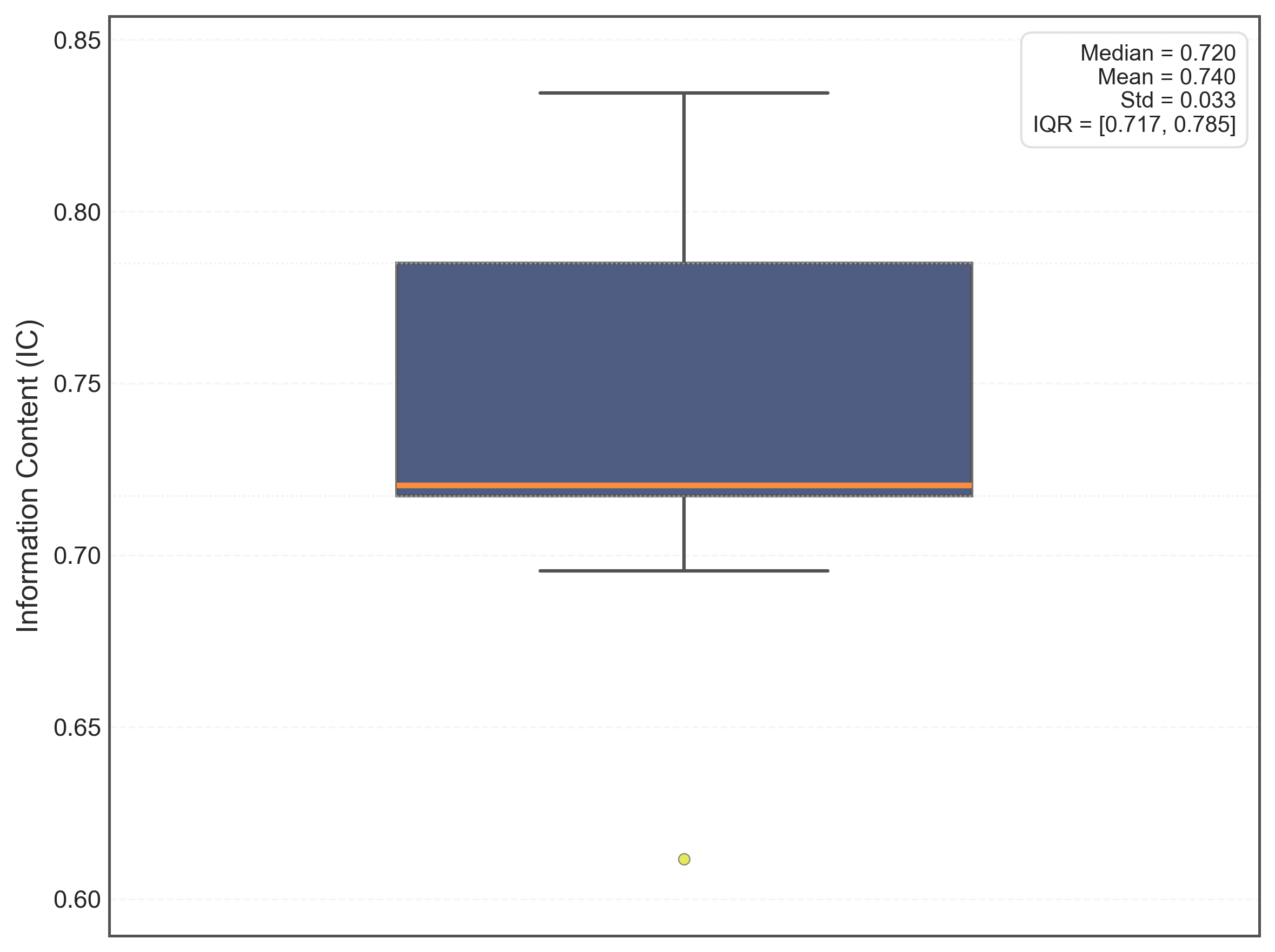}
    \caption{Distribution of IC values across all patient–gene explanations generated by ECKO. The box plot shows the median, mean, variability (IQR), and outliers, illustrating the robustness of explanatory paths provided for all 296 patients.}
    \label{fig:plot}
\end{figure}

We computed the distribution of IC scores across all patient–gene pairs (Figure~\ref{fig:plot}). The overall average IC was approximately 0.74, with an interquartile range between 0.717 and 0.785.

\subsection{Impact of ECKO}
ECKO is composed of 33 biomedical ontologies and datasets of multiple domains incorporated into a single resource that can be easily used and reused. While its current form has been shown to be enough to support personalized medicine solutions, it can also be tailored to future projects by incorporating new datasets and ontologies, ensuring that it is reusable and interoperable. As it comprehensively describes multiple biomedical domains, it can be successfully used in other projects with minimal alterations, functioning as a solid starting point for other applications. A tentative indication of the importance of ECKO is the nearly 500 downloads of its Zenodo repository, as of the writing of this article.

The KATY project use-case demonstrated that ECKO is capable of supporting AI solutions for personalized drug recommendation in ccRCC, solely using publicly available ontologies and datasets. Furthermore, it can successfully be used to generate explanations for the model predictions that are user-friendly and were validated by multiple focus groups of experts.

\section{Code Availability}
The code for the ontology alignment is available at \url{https://github.com/liseda-lab/holistic-matching-aml} and \url{https://github.com/liseda-lab/CMOM-RS}. The code for the data annotation is available at \url{https://github.com/liseda-lab/KATY-KG/}. The code for REx implementation is available at \url{https://github.com/liseda-lab/REx}.

\section{Author Contributions}
All authors contributed to reviewing the final manuscript. \textbf{Marta Silva: } Original draft, Methodology, Software, Validation, Data analysis, Visualization. \textbf{Daniel Faria: } Methodology, Supervision. \textbf{Laura Balbi: } Software, Validation, Data analysis, Data Curation. \textbf{Susana Nunes: } Methodology, Software, Validation, Data analysis, Visualization, Writing. \textbf{Aleksander Palkowski: } Resources, Data Curation. \textbf{Michal Waleron: } Resources, Data Curation. \textbf{Emilia Daghir-Wojtkowiak: } Resources, Data Curation. \textbf{Ashwin Adrian Kallor: } Resources, Data Curation. \textbf{Christophe Battail: } Data Curation. \textbf{Federico M. Corazza: } Software. \textbf{Manuel Fiorelli: } Methodology, Validation. \textbf{Armando Stellato: } Methodology, Validation. \textbf{Javier A. Alfaro: } Methodology, Resources, Data Curation. \textbf{Fabio M. Zanzotto: } Methodology, Supervision. \textbf{Catia Pesquita: } Methodology, Validation, Writing, Supervision.

\section{Competing Interests}
The authors declare no competing interests.

\paragraph{Acknowledgments} This work was supported by the KATY project, which has received funding from the European Union’s Horizon 2020 research and innovation program under grant agreement No 101017453. It was partially supported by FCT through the fellowships \\ https://doi.org/10.54499/2022.11895.BD (MS), https://doi.org/10.54499/2023.00653.BD (SN) and \\ https://doi.org/10.54499/2024.01208.BD (LB), and the LASIGE Research Unit, ref. UID/408/2025 - LASIGE. It was also partially supported by the CancerScan project by the EU's HORIZON Europe research and innovation programme under grant agreement No 101186829, and project 41, HfPT: Health from Portugal, funded by the Portuguese Plano de Recuperação e Resiliência.

\bibliographystyle{plain}
\bibliography{refs}
	
\end{document}